\documentclass[twocolumn,showpacs,preprintnumbers]{revtex4}
\usepackage{amsmath}
\usepackage{amssymb}
\usepackage{amsfonts}
\usepackage{graphicx}
\usepackage{dcolumn}
\usepackage{bm}

\input{tcilatex}

\begin{document}

\title{Common Markets, Strong Currencies \\
\& the Collective Welfare}
\author{Esteban Guevara Hidalgo$^{\dag \ddag }$}
\affiliation{$^{\dag }$Center for Nonlinear and Complex Systems, Universit\`{a} degli
Studi \\
dell'Insubria, Via Valleggio 11, 22100 Como, Italy \\
$^{\ddag }$SI\'{O}N, Autopista General Rumi\~{n}ahui, Urbanizaci\'{o}n Ed\'{e}%
n del Valle,\\
Sector 5, Calle 1 y Calle A \# 79, Quito, Ecuador}

\begin{abstract}
The so called \textquotedblleft globalization\textquotedblright\ process
(i.e. the inexorable integration of markets, currencies, nation-states,
technologies and the intensification of consciousness of the world as a
whole) has a behavior exactly equivalent to a system that is tending to a
maximum entropy state. This globalization process obeys a collective welfare
principle in where the maximum payoff is given by the equilibrium of the
system and its stability by the maximization of the welfare of the
collective besides the individual welfare. This let us predict the
apparition of big common markets and strong common currencies. They will
reach the \textquotedblleft equilibrium\textquotedblright\ by decreasing its
number until they reach a state characterized by only one common currency
and only one big common community around the world.

\end{abstract}

\pacs{03.65.-w, 02.50.Le, 03.67.-a, 89.65.Gh}
\maketitle
\email{esteban\_guevarah@yahoo.es}

\section{Introduction}

In recent works \cite{1,2,3,4,5,6,7} the relationships between quantum
mechanics and game theory were analyzed. Starting from the fact that a
physical system is modeled through quantum mechanics (and/or physics) and a
socioeconomical one through evolutionary game theory was shown how although
both systems are described through two theories apparently different both
are analogous and thus exactly equivalents.

That produced interesting results (like the quantum analogue of the
replicator dynamics, the quantization relationships for classical systems,
the quantum games entropies, the thermodynamical temperature of a
socioeconomical system, the collective welfare principle and the quantum
understanding of classical systems) and strange interesting consequences
like\ the analyzed in the following research.

The so called \textquotedblleft globalization\textquotedblright\ process
(i.e. the inexorable integration of markets, currencies, nation-states,
technologies and the intensification of consciousness of the world as a
whole) has a behavior exactly equivalent to a system that is tending to a
maximum entropy state.

This globalization process obeys a collective welfare principle in where the
maximum payoff is given by the equilibrium of the system and its stability
by the maximization of the welfare of the collective over the individual
welfare. This let us predict the apparition of big common markets and strong
common currencies. They will reach the \textquotedblleft
equilibrium\textquotedblright\ by decreasing its number until they get a
state characterized by only one common currency and only one big common
community around the world.

\section{Relationships between Quantum Mechanics \& Game Theory}

The clear resemblances and apparent differences between both theories and
between the properties both enjoy were a motivation to try to find an actual
relationship between both systems. In the next table we compare some
characteristic aspects of quantum mechanics and game theory.

{\scriptsize Table 1}

\begin{center}
\begin{tabular}{cc}
\hline
{\scriptsize Quantum Mechanics} & {\scriptsize Game Theory} \\ \hline
{\scriptsize n system members} & {\scriptsize n players} \\ 
{\scriptsize Quantum States} & {\scriptsize Strategies} \\ 
{\scriptsize Density Operator} & {\scriptsize Relative Frequencies Vector}
\\ 
{\scriptsize Von Neumann Equation} & {\scriptsize Replicator Dynamics} \\ 
{\scriptsize Von Neumann Entropy} & {\scriptsize Shannon Entropy} \\ 
{\scriptsize System Equilibrium} & {\scriptsize Payoff} \\ 
{\scriptsize Maximum Entropy} & {\scriptsize Maximum Payoff} \\ 
&  \\ \hline
\end{tabular}
\end{center}

To find an actual relationship between quantum mechanics and game theory
lets analyze more deeply both systems.

\subsection{Quantum Statistical Mechanics \\ \& the von Neumann Equation}

\subsubsection{Description of a Physical System}

A physical system is described through a Hilbert space where a state of that
system is described through a state vector (element of that Hilbert space) $%
\left\vert \Psi (t)\right\rangle $ which is postulated that contains all the
information of the system. An ensemble is a collection of identically
prepared physical systems. When each member of the ensemble is characterized
by the same state vector $\left\vert \Psi (t)\right\rangle $ it is called
pure ensemble. If each member has a probability $p_{i}$ of being in the
state $\left\vert \Psi _{i}(t)\right\rangle $ we have a mixed ensemble. Each
member of a mixed ensemble\ is a pure state and its evolution is given by
Schr\"{o}dinger equation. To describe correctly a statistical mixture of
states it is necessary the introduction of the density operator $\rho
(t)=\sum_{i=1}^{n}p_{i}\left\vert \Psi _{i}(t)\right\rangle \left\langle
\Psi _{i}(t)\right\vert $ which can be expressed in matrix form. The density
operator contains all the physically significant information we can obtain
about the ensemble in question. Any two ensembles that produce the same
density operator are physically indistinguishable. The diagonal elements $%
\rho _{nn}$ of \ the matrix which represents the density operator $\rho (t)$
represents the average probability of finding the system in the state $%
\left\vert n\right\rangle $ and its sum is equal to $1$. The non-diagonal
elements $\rho _{np}$ expresses the interference effects between the states $%
\left\vert n\right\rangle $ and $\left\vert p\right\rangle $ which can
appear when the state $\left\vert \Psi _{i}\right\rangle $ is a coherent
linear superposition of these states.

\subsubsection{Evolution of a Physical System}

The time evolution of the density operator is given by the von Neumann
equation%
\begin{equation}
i\hbar \frac{d\rho }{dt}=\left[ \hat{H},\rho \right] \text{,}  \label{1}
\end{equation}%
where $\hat{H}$ is the Hamiltonian of the physical system. The von Neumann
equation is only a generalization (and/or a matrix-operator representation) of the
Schr\"{o}dinger equation and the quantum analogue of Liouville's theorem
from classical statistical mechanics.

\subsection{Evolutionary Game Theory \\ \& the Replicator Dynamics}

\subsubsection{Description of a Socioeconomical System \\ \& the Notions of
Equilibrium}

Game theory \cite{8,9,10} is the study of decision making of competing
agents in some conflict situation. It tries to understand the birth and the
development of conflicting or cooperative behaviors among a group of
individuals who behave rationally and strategically according to their
personal interests. Each member in the group strive to maximize its welfare
by choosing the best courses of strategies from a cooperative or individual
point of view.

Evolutionary game theory \cite{11,12,13} does not rely on rational
assumptions but on the idea that the Darwinian process of natural selection 
\cite{14} drives organisms towards the optimization of reproductive success 
\cite{15}. Instead of working out the optimal strategy, the different
phenotypes in a population are associated with the basic strategies that are
shaped by trial and error by a process of natural selection or learning. The
natural selection process that determines how populations playing specific
strategies evolve is known as the replicator dynamics \cite{16,12,13,17}
whose stable fixed points are Nash Equilibria (NE) \cite{9}. The central
equilibrium concept of evolutionary game theory is the notion of
Evolutionary Stable Strategy (ESS) introduced by J. Smith and G. Price \cite%
{18,11}. An ESS is described as a strategy which has the property that if
all the members of a population adopt it, no mutant strategy could invade
the population under the influence of natural selection. ESS are interpreted
as stable results of processes of natural selection.

Quantum games have proposed a new point of view for the solution of the
classical problems and dilemmas in game theory. Quantum games are more
efficient than classical games and provide a saturated upper bound for this
efficiency \cite{19,20,21,22,23,24}.

A Nash equilibrium is a set of strategies, one for each player, such that no
player has an incentive to unilaterally change his action. Players are in
equilibrium if a change in strategies by any one of them would lead that
player to earn less than if he remained with his current strategy. A Nash
equilibrium satisfies the following condition%
\begin{equation}
E(p,p)\geq E(r,p)\text{,}  \label{2}
\end{equation}%
where $E(s_{i},s_{j})$ is a real number that represents the payoff obtained
by a player who plays the strategy $s_{i}$\ against an opponent who plays
the strategy $s_{j}$. A player can not increase his payoff if he decides to
play the strategy $r$ instead of $p.$

Consider a large population in which a two person game $G=(S,E)$ is played
by randomly matched pairs of animals generation after generation. Let $p$ be
the strategy played by the vast majority of the population, and let $r$ be
the strategy of a mutant present in small frequency. Both $p$ and $r$ can be
pure or mixed. An evolutionary stable strategy (ESS) $p$ of a symmetric
two-person game $G=(S,E)$ is a pure or mixed strategy for $G$ which
satisfies the following two conditions%
\begin{gather}
E(p,p)>E(r,p)\text{,}  \notag \\
\text{If }E(p,p)=E(r,p)\text{ then }E(p,r)>E(r,r)\text{.}  \label{3}
\end{gather}%
Since the stability condition only concerns to alternative best replies, $p$
is always evolutionarily stable if $(p,p)$ is a strict equilibrium point. An
ESS is also a Nash equilibrium since is the best reply to itself and the
game is symmetric. The set of all the strategies that are ESS is a subset of
the NE of the game. A population which plays an ESS can withstand an
invasion by a small group of mutants playing a different strategy. It means
that if a few individuals which play a different strategy are introduced
into a population in an ESS, the evolutionarily selection process would
eventually eliminate the invaders.

\subsubsection{Evolution of a Socioeconomical System}

Each agent in a n-player game where the $i^{th}$ player has as strategy
space $S_{i}$ is modeled by a population of players which have to be
partitioned into groups. Individuals in the same group would all play the
same strategy. Randomly we make play the members of the subpopulations
against each other. The subpopulations that perform the best will grow and
those that do not will shrink and eventually will vanish. The natural
selection process assures survival of the best players at the expense of the
others. The natural selection process that determines how populations
playing specific strategies evolve is known as the replicator dynamics%
\begin{equation}
\frac{dx_{i}}{dt}=\left[ f_{i}(x)-\left\langle f(x)\right\rangle \right]
x_{i}\text{.}  \label{4}
\end{equation}%
The probability of playing certain strategy or the relative frequency of
individuals using that strategy is denoted by frequency $x_{i}$. The fitness
function $f_{i}=\sum_{j=1}^{n}a_{ij}x_{j}$ specifies how successful each
subpopulation is, $\left\langle f(x)\right\rangle
=\sum_{k,l=1}^{n}a_{kl}x_{k}x_{l}$ is the average fitness of the population,
and $a_{ij}$ are the elements of the payoff matrix $A$%

\begin{equation}
\frac{dx_{i}}{dt}=\left[ \sum_{j=1}^{n}a_{ij}x_{j}-%
\sum_{k,l=1}^{n}a_{kl}x_{k}x_{l}\right] x_{i}\text{.}  \label{5}
\end{equation}%
The replicator dynamics rewards strategies that outperform the average by
increasing their frequency, and penalizes poorly performing strategies by
decreasing their frequency. The stable fixed points of the replicator
dynamics are Nash equilibria, it means that if a population reaches a state
which is a Nash equilibrium, it will remain there. It is important to note
that the replicator dynamics is a vectorial differential equation while von
Neumann equation can be represented in matrix form. If we would like to
compare both systems the first we would have to do is to try to compare
their evolution equations by trying to find a matrix representation of the
replicator dynamics and this is%
\begin{equation}
\frac{dX}{dt}=G+G^{T}\text{,}  \label{6}
\end{equation}%
where the relative frequencies matrix $X$ has as elements $x_{ij}=\left(
x_{i}x_{j}\right) ^{1/2}$ and%
\begin{eqnarray}
\left( G+G^{T}\right) _{ij} &=&\frac{1}{2}\sum_{k=1}^{n}a_{ik}x_{k}x_{ij} 
\notag \\
&&+\frac{1}{2}\sum_{k=1}^{n}a_{jk}x_{k}x_{ji}  \notag \\
&&-\sum_{k,l=1}^{n}a_{kl}x_{k}x_{l}x_{ij}  \label{7}
\end{eqnarray}%
are the elements of the matrix $\left( G+G^{T}\right) $. Moreover, from this
matrix representation we can find a Lax representation of the replicator
dynamics \cite{1}%
\begin{equation}
\frac{dX}{dt}=\left[ \left[ Q,X\right] ,X\right]  \label{8}
\end{equation}%
and with $\Lambda =\left[ Q,X\right] $%
\begin{equation}
\frac{dX}{dt}=\left[ \Lambda ,X\right] \text{.}  \label{9}
\end{equation}%
The matrix $\Lambda $ is equal to%
\begin{equation}
(\Lambda )_{ij}=\frac{1}{2}\left[ \left( \sum_{k=1}^{n}a_{ik}x_{k}\right)
x_{ij}-x_{ji}\left( \sum_{k=1}^{n}a_{jk}x_{k}\right) \right]  \label{10}
\end{equation}%
and $Q$ is a diagonal matrix which has as elements $q_{ii}=\frac{1}{2}%
\sum_{k=1}^{n}a_{ik}x_{k}$. If we take $\Theta =\left[ \Lambda ,X\right] $
equation (\ref{9}) becomes into $\frac{dX}{dt}=\Theta $, where the elements
of the matrix $\Theta $ are given by $\left( \Theta \right) _{ij}=\frac{1}{2}%
\sum_{k=1}^{n}a_{ik}x_{k}x_{ij}+\frac{1}{2}\sum_{k=1}^{n}a_{jk}x_{k}x_{ji}-%
\sum_{k,l=1}^{n}a_{lk}x_{k}x_{l}x_{ij}$. It is easy to realize that the
matrix commutative form of the replicator dynamics (\ref{9}) follows the
same dynamic than the von Neumann equation (\ref{1}). It can be shown that
the properties of their correspondent elements (matrixes) are similar, being
the properties corresponding to our quantum system more general than the
classical system.

\subsection{Actual Relationships between \newline
Quantum Mechanics \& Game Theory}

The following table shows some specific resemblances between quantum
statistical mechanics and evolutionary game theory.

{\scriptsize Table 2}

\begin{center}
\begin{tabular}{cc}
\hline
{\scriptsize Quantum Statistical Mechanics} & {\scriptsize Evolutionary Game
Theory} \\ \hline
{\scriptsize n system members} & {\scriptsize n population members} \\ 
{\scriptsize Each member in the state }$\left\vert \Psi _{k}\right\rangle $
& {\scriptsize Each member plays strategy }$s_{i}$ \\ 
$\left\vert \Psi _{k}\right\rangle $ {\scriptsize with} $p_{k}\rightarrow $
\ $\rho _{ij}${\scriptsize \ } & $s_{i}${\scriptsize \ }$\ \ \rightarrow $%
{\scriptsize \ }$\ \ x_{i}$ \\ 
$\rho ,$ $\ \ \sum_{i}\rho _{ii}{\scriptsize =1}$ & ${\scriptsize X,}$%
{\scriptsize \ \ }$\sum_{i}x_{i}{\scriptsize =1}$ \\ 
${\scriptsize i\hbar }\frac{d\rho }{dt}{\scriptsize =}\left[ \hat{H},\rho %
\right] $ & $\frac{dX}{dt}{\scriptsize =}\left[ \Lambda ,X\right] $ \\ 
${\scriptsize S=-Tr}\left\{ {\scriptsize \rho }\ln {\scriptsize \rho }%
\right\} $ & ${\scriptsize H=-}\sum_{i}{\scriptsize x}_{i}\ln 
{\scriptsize x}_{i}$ \\ 
&  \\ \hline
\end{tabular}
\end{center}

Although a physical system is modeled and described mathematically through
quantum mechanics while a socioeconomical is modeled through game theory
both systems seem to have a similar behavior. Both are composed by $n$
members (particles, subsystems, players, states, etc.). Each member of our
systems is described by a state or a strategy which has assigned a
determined probability. The quantum mechanical system is described by a
density operator $\rho $ whose elements represent the system average
probability of being in a determined state. The socioeconomical system is
described through a relative frequencies matrix $X$ whose elements represent
the frequency of players playing a determined strategy. The evolution
equation of the relative frequencies matrix $X$ (which describes our
socioeconomical system) is given by a Lax form of the replicator dynamics
which was shown that follows the same dynamic than the evolution equation of
the density operator (i.e. the von Neumann equation).

\section{Direct Consequences in the Analogous Behavior of Quantum Mechanics \& Game Theory}

\subsection{Quantization Relationships}

We can propose the next \textquotedblleft quantization
relationships\textquotedblright %

\begin{gather}
x_{i}\rightarrow \sum_{k=1}^{n}\left\langle i\left\vert \Psi _{k}\right.
\right\rangle p_{k}\left\langle \Psi _{k}\left\vert i\right. \right\rangle
=\rho _{ii}\text{,}  \notag \\
(x_{i}x_{j})^{1/2}\rightarrow \sum_{k=1}^{n}\left\langle i\left\vert \Psi
_{k}\right. \right\rangle p_{k}\left\langle \Psi _{k}\left\vert j\right.
\right\rangle =\rho _{ij}\text{.}  \label{13}
\end{gather}%

A population will be represented by a quantum system in which each
subpopulation playing strategy $s_{i}$ will be represented by a pure
ensemble in the state $\left\vert \Psi _{k}(t)\right\rangle $ and with
probability $p_{k}$. The probability $x_{i}$ of playing strategy $s_{i}$ or
the relative frequency of the individuals using strategy $s_{i}$ in that
population will be represented as the probability $\rho _{ii}$ of finding
each pure ensemble in the state $\left\vert i\right\rangle $ \cite{1}.

\subsection{Quantum Replicator Dynamics}

Through the last quantization relationships the replicator dynamics (in
matrix commutative form) takes the form of the equation of evolution of
mixed states i.e. the von Neumann equation is the quantum analogue of \ the
replicator dynamics. And also $X\longrightarrow \rho $, $\Lambda
\longrightarrow -\frac{i}{\hbar }\hat{H}$, and $H(x)\longrightarrow S(\rho )$
\cite{1,3}.

\subsection{Quantum Games Entropy}

Classically, the entropy of our system is given by%
\begin{equation}
H=-Tr\left\{ X\ln X\right\} \text{.}  \label{14}
\end{equation}%
When the non diagonal elements of matrix $X$ are equal to zero it turns to
the Shannon entropy over the elements of the relative frequency vector $x$,
i.e. $H=-\sum_{i=1}^{n}x_{i}\ln x_{i}$. By supposing that the vector of
relative frequencies $x(t)$ evolves in time following the replicator
dynamics\ (\ref{9}) the evolution of the entropy of our system would be
given by \cite{3}%
\begin{equation}
\frac{dH}{dt}=Tr\left\{ U(\tilde{H}-X)\right\} \text{,}  \label{15}
\end{equation}%
where $U_{i}=\left[ f_{i}(x)-\left\langle f(x)\right\rangle \right] $, and $%
\tilde{H}$ comes from $H=Tr\tilde{H}$.

The entropy of a quantum system is given by the von Neumann entropy%
\begin{equation}
S(t)=-Tr\left\{ \rho \ln \rho \right\}  \label{16}
\end{equation}%
which in a far from equilibrium system also vary in time until it reaches
its maximum value. When the dynamics is chaotic the variation with time of
the physical entropy goes through three successive, roughly separated stages 
\cite{25}. In the first one, $S(t)$ is dependent on the details of the
dynamical system and of the initial distribution, and no generic statement
can be made. In the second stage, $S(t)$ is a linear increasing function of
time ($\frac{dS}{dt}=const.$). In the third stage, $S(t)$ tends
asymptotically towards the constant value which characterizes equilibrium ($%
\frac{dS}{dt}=0$). With the purpose of calculating the time evolution of
entropy we approximated the logarithm of $\rho $ by series i.e. $\ln \rho
=(\rho -I)-\frac{1}{2}(\rho -I)^{2}+\frac{1}{3}(\rho -I)^{3}$... and \cite{3}%
\begin{eqnarray}
\frac{dS(t)}{dt} &=&\frac{11}{6}\sum\limits_{i}\frac{d\rho _{ii}}{dt} 
\notag \\
&&-6\sum\limits_{i,j}\rho _{ij}\frac{d\rho _{ji}}{dt}  \notag \\
&&+\frac{9}{2}\sum\limits_{i,j,k}\rho _{ij}\rho _{jk}\frac{d\rho _{ki}}{dt}
\notag \\
&&-\frac{4}{3}\sum\limits_{i,j,k,l}\rho _{ij}\rho _{jk}\rho _{kl}\frac{%
d\rho _{li}}{dt}+\zeta \text{.}  \label{17}
\end{eqnarray}%

\subsection{Games Analysis from Quantum \newline
Information Theory}

If we define an Entropy \cite{2,3,26,27,28,29,30} over a random variable $%
S^{A}$ (player's $A$ strategic space) which can take the values $s_{i}^{A}$
with the respective probabilities $x_{i}^{A}$ i.e. $H(A)\equiv
-\sum_{i=1}^{n}x_{i}\log _{2}x_{i}$, we could interpret the entropy of our
game as a measure of uncertainty before we learn what strategy player $A$ is
going to use \cite{3}. If we do not know what strategy a player is going to
use every strategy becomes equally probable and our uncertainty becomes
maximum and it is greater while greater is the number of strategies. If we
would know the relative frequency with which player $A$ uses any strategy we
can prepare our reply in function of the most probable player $A$ strategy.
That would be our actual best reply which in that moment would let us
maximize our payoff due to our uncertainty. Obviously our uncertainty vanish
if we are sure about the strategy our opponent is going to use. The complete
knowledge of the rules of a game and the reserve in our strategies becomes
an advantage over an opponent who does not know the game rules or who always
plays in a same predictive way. To become a game fair, an external referee
should make the players to know completely the game rules and the strategies
that the players can use.

If player $B$ decides to play strategy $s_{j}^{B}$ against player $A$
(who plays strategy $s_{i}^{A})$ our total uncertainty about the pair $(A,B)$
can be measured by an external \textquotedblleft referee\textquotedblright\
through the joint entropy of the system $H(A,B)\equiv -\sum_{i,j}x_{ij}\log
_{2}x_{ij}$, $x_{ij}$ is the joint probability to find $A$ in state $s_{i}$
and $B$ in state $s_{j}$. This is smaller or at least equal than the sum of
the uncertainty about $A$ and the uncertainty about $B,$ $H(A,B)\leq
H(A)+H(B)$. The interaction and the correlation between $A$ and $B$ reduces
the uncertainty due to the sharing of information. There can be more
predictability in the whole than in the sum of the parts. The uncertainty
decreases while more systems interact jointly creating a new only system.

We can measure how much information $A$ and $B$ share and have an idea of
how their strategies or states are correlated by their mutual or correlation
entropy $H(A:B)\equiv -\sum_{i,j}x_{ij}\log _{2}x_{i:j}$, with $x_{i:j}=%
\frac{\sum_{i}x_{ij}\sum_{j}x_{ij}}{x_{ij}}$. It can be seen easily as $%
H(A:B)\equiv H(A)+H(B)-H(A,B)$. The joint entropy would equal the sum of
each of $A$'s and $B$'s entropies only in the case that there are no
correlations between $A$'s and $B$'s states. In that case, the mutual
entropy vanishes and we could not make any predictions about $A$ just from
knowing something about $B$.

If we know that $B$ decides to play strategy $s_{j}^{B}$ we can determinate
the uncertainty about $A$ through the conditional entropy $H(A\mid B)\equiv
H(A,B)-H(B)=-\sum_{i,j}x_{ij}\log _{2}x_{i\mid j}$ with $x_{i\mid j}=\frac{%
x_{ij}}{\sum_{i}x_{ij}}$. If this uncertainty is bigger or equal to zero
then the uncertainty about the whole is smaller or at least equal than the
uncertainty about $A$, i.e. $H(A:B)\leq H(A)$. Our uncertainty about the
decisions of player $A$ knowing how $B$ and $C$ plays is smaller or at least
equal than our uncertainty about the decisions of $A$ knowing only how $B$
plays $H(A\mid B,C)\leq H(A\mid B)$ i.e. conditioning reduces entropy. If
the behavior of the players of a game follows a Markov chain i.e. $%
A\rightarrow B\rightarrow C$ then $H(A)\geq H(A:B)\geq H(A:C)$ i.e. the
information can only reduces in time. Also any information $C$ shares with $%
A $ must be information which $C$ also shares with $B$, $H(C:B)\geq H(C:A)$.

Two external observers of the same game can measure the difference in their
perceptions about certain strategy space of the same player $A$ by its
relative entropy. Each of them could define a relative frequency vector, $x$
and $y$, and the relative entropy over these two probability distributions
is a measure of its closeness $H(x\parallel y)\equiv \sum_{i}x_{i}\log
_{2}x_{i}-\sum_{i}x_{i}\log _{2}y_{i}$. We could also suppose that $A$ could
be in two possible states i.e. we know that $A$ can play of two specific but
different ways and each way has its probability distribution (again $x$ and $%
y$ that also is known). Suppose that this situation is repeated exactly $N$
times or by $N$ people. We can made certain \textquotedblleft
measure\textquotedblright , experiment or \textquotedblleft
trick\textquotedblright\ to determine which the state of the player is. The
probability that these two states can be confused is given by the classical
or the quantum Sanov's theorem \cite{27,31,32,33}.

By analogy with the Shannon entropies it is possible to define conditional,
mutual and relative quantum entropies which also satisfy many other
interesting properties that do not satisfy their classical analogues. For
example, the conditional entropy $S(A\mid B)$ can be negative and its
negativity always indicates that two systems (in this case players) are
entangled and indeed, how negative the conditional entropy is provides a
lower bound on how entangled the two systems are \cite{34}. If $\lambda _{i}$
are the eigenvalues of $\rho $ then von Neumann's definition can be
expressed as $S(\lambda )=-\sum_{i}\lambda _{i}\ln \lambda _{i}$ and it
reduces to a Shannon entropy if $\rho $ is a mixed state composed of
orthogonal quantum states \cite{35}. Our uncertainty about the mixture of
states $S(\sum_{i}p_{i}\rho _{i})$ should be higher than the average
uncertainty of the states $\sum_{i}p_{i}S(\rho _{i})$.

\subsection{Thermodynamical Temperature \newline
of a Socioeconomical System}

By other hand, in statistical mechanics entropy can be regarded as a
quantitative measure of disorder. It takes its maximum possible value $\ln n$
in a completely random ensemble in which all quantum mechanical states are
equally likely and is equal to zero if $\rho $ is pure i.e. when all its
members are characterized by the same quantum mechanical state ket. Entropy
can be maximized subject to different constraints. Generally, the result is
a probability distribution function. We maximize $S(\rho )$ subject to the
constraints $\delta Tr\left( \rho \right) =0$ and $\delta \left\langle
E\right\rangle =0$ and the result is%
\begin{equation}
\rho _{ii}=\frac{e^{-\beta E_{i}}}{\sum_{k}e^{-\beta E_{k}}}  \label{18}
\end{equation}%
which is the condition that the density operator must satisfy to our system
tends to maximize its entropy $S$. Without the internal energy constraint $%
\delta \left\langle E\right\rangle =0$ we obtain $\rho _{ii}=\frac{1}{N}$
which is the $\beta \rightarrow 0$\ limit (\textquotedblleft high -
temperature limit\textquotedblright ) in equation (\ref{18}) in where a
canonical ensemble becomes a completely random ensemble in which all energy
eigenstates are equally populated. In the opposite low - temperature limit $%
\beta \rightarrow \infty $ tell us that a canonical ensemble becomes a pure
ensemble where only the ground state is populated \cite{36}. The parameter $%
\beta $ is related inversely to the \textquotedblleft
temperature\textquotedblright\ $\tau $ of the system, $\beta =\frac{1}{\tau }
$. We can rewrite entropy in function of the partition function $%
Z=\sum_{k}e^{-\beta E_{k}}$, $\beta $ and $\left\langle E\right\rangle $ via 
$S=\ln Z+\beta \left\langle E\right\rangle $. From the partition function we
can know some parameters that define the system like $\left\langle
E\right\rangle $ and $\left\langle \Delta E^{2}\right\rangle $. We can also
analyze the variation of entropy with respect to the average energy of the
system%
\begin{gather}
\frac{\partial S}{\partial \left\langle E\right\rangle }=\frac{1}{\tau }%
\text{,}  \label{19} \\
\frac{\partial ^{2}S}{\partial \left\langle E\right\rangle ^{2}}=-\frac{1}{%
\tau ^{2}}\frac{\partial \tau }{\partial \left\langle E\right\rangle }
\label{20}
\end{gather}%
and with respect to the parameter $\beta $%
\begin{gather}
\frac{\partial S}{\partial \beta }=-\beta \left\langle \Delta
E^{2}\right\rangle \text{,}  \label{21} \\
\frac{\partial ^{2}S}{\partial \beta ^{2}}=\frac{\partial \left\langle
E\right\rangle }{\partial \beta }+\beta \frac{\partial ^{2}\left\langle
E\right\rangle }{\partial \beta ^{2}}\text{.}  \label{22}
\end{gather}%

\subsection{The Applicability of Physics to Economics}

Although both systems analyzed are described through two theories apparently
different both are analogous and thus exactly equivalents. So, we could make
use of some of the concepts, laws and definitions in physics for the best
understanding of the behavior of economics and biology. Quantum mechanics
could be a much more general theory that we had thought. It could encloses
theories like games and evolutionary dynamics. From this point of view many
of the equations, concepts and its properties defined quantically must be
more general that its classical analogues.

It is important to remember that we are dealing with very general and
unspecific terms, definitions, and concepts like state, game and system. Due
to this, the theories that have been developed around these terms like
quantum mechanics, statistical physics, information theories and game
theories enjoy of this generality quality and could be applicable to model
any system depending on what we want to mean for game, state, or system.
Objectively these words can be and represent anything. Once we have defined
what system is in our model, we could try to understand what kind of
\textquotedblleft game\textquotedblright\ is developing between its members
and how they accommodate their \textquotedblleft states\textquotedblright\
in order to get their objectives. This would let us visualize what
temperature, energy and entropy would represent in our specific system
through the relationships, properties and laws that were defined before when
we described a physical system \cite{4,5,6}.

\subsection{The Collective Welfare Principle \& the \\ Quantum Understanding 
of Classical Systems}

If our systems are analogous and thus exactly equivalents, our physical
equilibrium (maximum entropy) should be also exactly equivalent to our
socioeconomical equilibrium (NE or ESS). And if the natural trend of a
physical system is to a maximum entropy state, should not a socioeconomical
system trend be also to a maximum entropy state which would have to be its
state of equilibrium? Has a socioeconomical system something like a
\textquotedblleft natural trend\textquotedblright ?

Based specially on the analogous behavior between quantum mechanics and game
theory\textbf{,} it is suggested the\textbf{\ }following (quantum)\textbf{\ }%
understanding of our (classical and/or socioeconomical) system: If in an
isolated system each of its accessible states do not have the same
probability, the system is not in equilibrium. The system will vary and will
evolve in time until it reaches the equilibrium state in where the
probability of finding the system in each of the accessible states is the
same. The system will find its more probable configuration in which the
number of accessible states is maximum and equally probable. The whole
system will vary and rearrange its state and the states of its ensembles
with the purpose of maximize its entropy and reach its equilibrium state. We
could say that the purpose and maximum payoff of a physical system is its
maximum entropy state. The system and its members will vary and rearrange
themselves to reach the best possible state for each of them which is also
the best possible state for the whole system.

This can be seen like a microscopical cooperation between quantum objects to
improve their states with the purpose of reaching or maintaining the
equilibrium of the system. All the members of our quantum system will play a
game in which its maximum payoff is the equilibrium of the system. The
members of the system act as a whole besides individuals like they obey a
rule in where they prefer the welfare of the collective over the welfare of
the individual. This equilibrium is represented in the maximum entropy of
the system in where the system resources are fairly distributed over its
members. The system is stable only if it maximizes the welfare of the
collective above the welfare of the individual. If it is maximized the
welfare of the individual above the welfare of the collective the system
gets unstable and eventually it collapses (Collective Welfare Principle) 
\cite{1,3,4,5,6,7}.

\section{The Equilibrium Process called Globalization}

Lets discuss how the world process that it is called \textquotedblleft
globalization\textquotedblright\ has a behavior exactly equivalent to a
system that is tending to a maximum entropy state.

\subsection{Globalization}

Globalization represents the inexorable integration of markets,
nation-states, currencies, technologies \cite{37} and the intensification of
consciousness of the world as a whole \cite{38}.\textbf{\ }This refers to an
increasing global connectivity, integration and interdependence in the
economic, social, technological, cultural, political, and ecological spheres 
\cite{39}. Globalization has various aspects which affect the world in
several different ways such as \cite{39} the emergence of worldwide
production markets and broader access to a range of goods for consumers and
companies (industrial), the emergence of worldwide financial markets and
better access to external financing for corporate, national and subnational
borrowers (financial), the realization of a global common market, based on
the freedom of exchange of goods and capital (economical), the creation of a
world government which regulates the relationships among nations and
guarantees the rights arising from social and economic globalization
(political) \cite{40}, the increase in information flows between
geographically remote locations (informational), the growth of
cross-cultural contacts (cultural), the advent of global environmental
challenges that can not be solved without international cooperation, such as
climate change, cross-boundary water and air pollution, over-fishing of the
ocean, and the spread of invasive species (ecological) and the achievement
of free circulation by people of all nations (social).

\subsubsection{Economical Globalization}

In economics, globalization is the convergence of prices, products, wages,
rates of interest and profits towards developed country norms \cite{41}.
Globalization of the economy depends on the role of human migration,
international trade, movement of capital, and integration of financial
markets. Economic globalization can be measured around the four main
economic flows that characterize globalization such as goods and services
(e.g. exports plus imports as a proportion of national income or per capita
of population), labor/people (e.g. net migration rates; inward or outward
migration flows, weighted by population), capital (e.g. inward or outward
direct investment as a proportion of national income or per head of
population), and technology. To what extent a nation-state or culture is
globalized in a particular year has until most recently been measured
employing simple proxies like flows of trade, migration, or foreign direct
investment, as described above. A multivariate approach to measuring
globalization is the recent index calculated by the Swiss Think tank KOF 
\cite{42}. The index measures the three main dimensions of globalization:
economic, social, and political.

\subsubsection{Big Communities \& Strong Currencies}

Maybe the firsts of these so called states-nations, communities,
\textquotedblleft unions\textquotedblright , common markets, etc. were the
Unites States of America and the USSR (now the Russian Federation). Both
consists in a set or group of different nations or states under the same
basic laws or principles (constitution), policies, objectives and an economy
characterized by a same currency. Although each state or nation is a part of
a big community each of them can take its own decisions and choose its own
way of government, policies, laws and punishments (e.g. death penalty) but
subject to a constitution (which is no more than a \textquotedblleft common
agreement\textquotedblright )\ and also subject to the decisions of the
\textquotedblleft congress\textquotedblright\ of the community which
regulates the whole and the decisions of the parts. The United
States of America consists in 50 states and a federal district. It also has
many dependent territories located in the Antilles and Oceania. The currency
in The United States is the\ Dollar. The Russian Federation consists in a
big number of political subdivisions (88 components). There are 21 republics
inside the federation with a big degree of autonomy over most of the
aspects. The rest of territory consists in 48 provinces known as \'{o}blast
and six regions (kray), between which there are 10 autonomic districts and
an autonomic \'{o}blast and 2 federal cities (Moscow and San Petersburg).
Recently, seven federal districts have been added. The currency in Russia is
the\ Rublo \cite{39}.

The European Union stands as an example that the world should emulate by its
sharing rights, responsibilities, and values,\ including the obligation to
help the less fortunate. The most fundamental of these values is democracy,
understood to entail not merely periodic elections, but also active and
meaningful participation in decision making, which requires an engaged civil
society, strong freedom of information norms, and a vibrant and diversified
media that are not controlled by the state or a few oligarchs. The second
value is social justice. An economic and political system is to\ be judged
by the extent to which individuals are able to flourish and realize their
potential. As individuals, they are part of an ever-widening circle of
communities, and they can realize their potential only if they live in
harmony with each other. This, in turn, requires a sense of responsibility
and solidarity \cite{43}.

The meeting of 16 national leaders at the second East Asia Summit (EAS) on
the Philippine island of Cebu in January 2007 offered the promise of the
politically fractious but economically powerful Asian mega-region one day
coalescing into a single meaningful unit \cite{44}.

Seth Kaplan has offered the innovative idea of a West African Union (the 15
West African countries stretching from Senegal to Nigeria) to help solve West
Africa's deep-rooted problems \cite{45}.

In South America has been proposed the creation of a Latin-American
Community which is an offer for the integration, the struggle against the
poverty and the social exclusion of the countries of Latin-America. It is
based on the creation of mechanisms to create cooperative advantages between
countries that let balance the asymmetries between the countries of the
hemisphere and the cooperation of funds to correct the inequalities of the
weak countries against the powerful nations. The economy ministers of
Paraguay, Brazil, Argentina, Ecuador, Venezuela and Bolivia agreed in the
\textquotedblleft Declaraci\'{o}n de Asunci\'{o}n\textquotedblright\ to
create the Bank of the South and invite the rest of countries to add to this
project. The Brazilian economy minister Mantega stand out that the new bank
is going to consolidate the economic, social and politic block that is
appearing in South America and now they have to point to the creation of a
common currency. Recently, Uruguay has also accepted the offer of the
creation of the bank and the common currency and is expected that more
countries add to this offer \cite{46}.

\subsection{The Equilibrium Process}

After analyzing our systems we concluded that a socioeconomical system has a
behavior exactly equivalent that a physical system. Both systems evolve in
analogous ways and to analogous states. A system where its members are in
Nash Equilibrium (or ESS) is exactly equivalent to a system in a maximum
entropy state. The stability of the system is based on the maximization of
the welfare of the collective above the welfare of the individual. The
natural trend of a physical system is to a maximum entropy state, should not
a socioeconomical system trend be also to a maximum entropy state which
would have to be its state of equilibrium? Has a socioeconomical system
something like a \textquotedblleft natural trend\textquotedblright ?

From our analysis a population can be represented by a quantum system in
which each subpopulation playing strategy $s_{i}$ will be represented by a
pure ensemble in the state $\left\vert \Psi _{k}(t)\right\rangle $ and with
probability $p_{k}$. The probability $x_{i}$ of playing strategy $s_{i}$ or
the relative frequency of the individuals using strategy $s_{i}$ in that
population will be represented as the probability $\rho _{ii}$ of finding
each pure ensemble in the state $\left\vert i\right\rangle $. Through these
quantization relationships the replicator dynamics (in matrix commutative
form) takes the form of the equation of evolution of mixed states i.e. the
von Neumann equation is the quantum analogue of \ the replicator dynamics.

Our now \textquotedblleft quantum statistical\textquotedblright\ system
composed by quantum objects represented by quantum states which represent
the strategies with which \textquotedblleft players\textquotedblright\
interact is characterized by certain interesting physical parameters like
temperature, entropy and energy and has a similar or analogous behavior.

In this statistical mixture of ensembles (each ensemble is characterized by
a state and each state has assigned a determined probability) its natural
trend is to its maximum entropy state. If each of its accessible states do
not have the same probability, the system will vary and will evolve in time
until it reaches the equilibrium state in where the probability of finding
the system in each of the accessible states is the same and its number is
maximum. In this equilibrium state or maximum entropy state the system
\textquotedblleft resources\textquotedblright\ are fairly distributed over
its members. Each ensemble will be equally probable, will be characterized
by a same temperature and in a stable state.

Socioeconomically and based on our analysis, our world could be understood
as a statistical mixture of \textquotedblleft ensembles\textquotedblright\
(countries for example). Each of these ensembles are characterized by a
determined state and a determined probability. But more important, each
\textquotedblleft country\textquotedblright\ is characterized by a specific
\textquotedblleft temperature\textquotedblright\ which is a measure\textbf{\ 
}of the socioeconomical activity of that ensemble. That temperature is
related with the activity or with the interactions between the members of
the ensemble. The system will evolve naturally to a maximum entropy state.
Each pure ensemble of this statistical mixture will vary and accommodate its
state until get the \textquotedblleft thermal equilibrium\textquotedblright\
first with its nearest neighbors creating new big ensembles characterized
each of them by a same temperature. Then with the time, these new big
ensembles will seek its \textquotedblleft thermal
equilibrium\textquotedblright\ between themselves and with its nearest
neighbors\ creating new bigger ensembles. The system will continue evolving
naturally in time until the whole system get an only state characterized by
a same \textquotedblleft temperature\textquotedblright .

This behavior is very similar to what has been called globalization. The
process of equilibrium that is absolutely equivalent to a system that is
tending to a maximum entropy state is the actual globalization. This
analysis predicts the apparition of big common \textquotedblleft
markets\textquotedblright\ or (economical, political, social, etc.)
communities of countries (European Union, Asian Union, Latin-American
Community, African Union, Mideast Community, Russia and USA) and strong
common currencies (dollar, euro, yen, sol, etc.). The little and poor
economies eventually will be unavoidably absorbed by these \textquotedblleft
markets\textquotedblright\ and these currencies. If this process continues
these markets or communities will find its \textquotedblleft
equilibrium\textquotedblright\ by decreasing its number until reach a state
in where there exists only one big common community (or market) and only one
common currency around the world.

\section{Conclusions}

Although both systems analyzed are described through two theories apparently
different (quantum mechanics and game theory) both are analogous and thus
exactly equivalents. A socioeconomical system has a behavior exactly
equivalent that a physical system. Both systems evolve in analogous ways and
to analogous points. The quantum analogue of the replicator dynamics is the
von Neumann equation. A system where its members are in Nash Equilibrium (or
ESS) is exactly equivalent to a system in a maximum entropy state. The
natural trend of both systems is to its maximum entropy state which is its
state of equilibrium.

The so called \textquotedblleft globalization\textquotedblright\ process
(i.e. the inexorable integration of markets, currencies, nation-states,
technologies and the intensification of consciousness of the world as a
whole) has a behavior exactly equivalent to a system that is tending to a
maximum entropy state. This globalization process obeys a collective welfare
principle in where the maximum payoff is given by the equilibrium of the
system and its stability by the maximization of the welfare of the
collective over the individual welfare. This let us predict the apparition
of big common markets and strong common currencies that will reach the
\textquotedblleft equilibrium\textquotedblright\ by decreasing its number
until they get a state characterized by only one common currency and only
one big common community around the world.

\begin{acknowledgments}
The author (Esteban Guevara) undertook this work with the support of the
Abdus Salam International Centre for Theoretical Physics ICTP at Trieste,
Italy and its Programme for Training and Research in Italian Laboratories
ICTP-TRIL which let him continue with his research at the Center for
Nonlinear and Complex Systems at Universit\`{a} degli Studi dell'Insubria in
Como.
\end{acknowledgments}


\begin{thebibliography}{99}

\bibitem{1} E. Guevara H., \textit{Quantum Replicator Dynamics}, Physica A 
\textbf{369/2}, 393-407 (2006).

\bibitem{2} E. Guevara H., \textit{Introduction to the study of Entropy in
Quantum Games}, quant-ph/0604170.

\bibitem{3} E. Guevara H., \textit{Quantum Games Entropy}, Physica A \textbf{%
383/2}, 797-804 (2007).

\bibitem{4} E. Guevara H., \textit{The Why of the applicability of
Statistical Physics to Economics}, physics/0609088.

\bibitem{5} E. Guevara H., \textit{Quantum Econophysics}, in Proceedings of
Quantum Interaction 2007, AAAI Spring Symposia Series, Stanford University,
Palo Alto, published by the American Association of Artificial Intelligence,
AAAI PRESS Technical Report\textbf{\ SS-07-08}, 158-165 (2007).

\bibitem{6} E. Guevara H., \textit{EGT through Quantum Mechanics \& from
Statistical Physics to Economics}, arXiv:0705.0029v1.

\bibitem{7} E. Guevara H., \textit{Maximum Entropy, the Collective Welfare
Principle and the Globalization Process}, arXiv:0707.1897v1

\bibitem{8} J. von Neumann and O. Morgenstern, \textit{The Theory of Games
and Economic Behavior} (Princeton \ University Press, Princeton, 1947).

\bibitem{9} R. B. Myerson, \textit{Game Theory: An Analysis of Conflict}
(MIT Press, Cambridge, 1991).

\bibitem{10} M. A. Nowak and K. Sigmund, Nature \textbf{398}, 367 (1999).

\bibitem{11} J. M. Smith, \textit{Evolution and The Theory of Games}
(Cambridge University Press, Cambridge, UK, 1982).

\bibitem{12} J. Hofbauer and K. Sigmund, \textit{Evolutionary Games and
Replicator Dynamics} (Cambridge University Press, Cambridge, UK, 1998).

\bibitem{13} J. Weibul, \textit{Evolutionary Game Theory} (MIT Press,
Cambridge, MA, 1995).

\bibitem{14} R. A. Fisher, \textit{The Genetic Theory of Natural Selection}
(Oxford, Clarendon Press, 1930).

\bibitem{15} P. Hammerstein and R. Selten, \textit{Game Theory and
Evolutionary Biology} (Handbook of Game Theory. Vol 2. Elsevier B.V., 1994).

\bibitem{16} P. D. Taylor and L. B. Jonker , \textit{Evolutionary stable
strategies and game dynamics}, Mathematical Biosciences \textbf{40}, 145-156
(1978).

\bibitem{17} R. Cressman, \textit{The Stability Concept of Evolutionary Game
Theory: A Dynamic Approach} (Springer-Verlag, New York, 1992).

\bibitem{18} J. M. Smith and G. R. Price, \textit{The logic of animal
conflict}, Nature \textbf{246}, 15 (1973).

\bibitem{19} D. A. Meyer, Phys. Rev. Lett. \textbf{82}, 1052-1055 (1999).

\bibitem{20} J. Eisert, M. Wilkens and M. Lewenstein, Phys. Rev. Lett. 
\textbf{83}, 3077 (1999).

\bibitem{21} L. Marinatto and T. Weber, Phys. Lett. A \textbf{272}, 291
(2000).

\bibitem{22} A. P. Flitney and D. Abbott, Proc. R. Soc. (London) A \textbf{%
459}, 2463-74 (2003).

\bibitem{23} E. W. Piotrowski and J. Sladkowski, Int. J. Theor. Phys. 
\textbf{42}, 1089 (2003).

\bibitem{24} Azhar Iqbal, PhD thesis, quant-ph/0503176.

\bibitem{25} M. Baranger, V. Latora and A. Rapisarda, cond-mat/0007302.

\bibitem{26} C. Shannon, \textit{A mathematical theory of communication},
Bell System Tech. Jour. \textbf{27}, 379-423 (1948).

\bibitem{27} T. M. Cover and J. A. Thomas, \textit{Elements of Information
Theory} (Wiley, New York, 1991).

\bibitem{28} R. Landauer, Phys. Today \textbf{44}, 23-29 (1991).

\bibitem{29} R. Gray, \textit{Entropy and Information Theory}
(Springer-Verlag, New York, 1990).

\bibitem{30} A. Wehrl, Rev. Mod. Phys. \textbf{50}, 221-260 (1978).

\bibitem{31} B. Schumacher and M. Westmoreland, quant-ph/0004045.

\bibitem{32} F. Hiai and D. Petz, Comm. Math. Phys. \textbf{143}, 99 (1991).

\bibitem{33} V. Vedral, M. B. Plenio, K. Jacobs, and P. L. Knight, Phys.
Rev. A \textbf{56}, 4452 (1997).

\bibitem{34} M. A. Nielsen and I. L. Chuang, \textit{Quantum Computation and
Quantum Information} (Cambridge University Press, Cambridge, 2000).

\bibitem{35} N. J. Cerf and C. Adami, Phys. Rev. Lett. \textbf{79}, 5194
(1997).

\bibitem{36} J. J. Sakurai, \textit{Modern Quantum Mechanics} (Addison -
Wesley, 1994).

\bibitem{37} T.L. Friedman, \textit{The Lexus and the Olive Tree}, p.7-8, 1999.

\bibitem{38} R. Robertson, \textit{Globalization},p.8 , 1992.

\bibitem{39} Wikipedia.

\bibitem{40} M. Albrow, \textit{The Global Age},p.88 , 1996.

\bibitem{41} I. Shariff, \textit{Global Economic integration: Prospects and
Problems}, International Journal of Development Economics. Development
Review, \textbf{Vol1}, No.2, 163-178 (2003).

\bibitem{42} KOF Index of Globalization (http://www.globalization-index.org).

\bibitem{43} Joseph Stiglitz, \textit{The European Union's Global Mission},
Joseph Stiglitz talks of Europe's achievements and future challenges. Date
Posted on Global Envision: May 14, 2007.

\bibitem{44} Bennett Richardson, \textit{An East Asian Community? Not So Fast%
}. Date Posted on Global Envision: May 02, 2007.

\bibitem{45} Seth Kaplan, \textit{A West African Union}. Date Posted on
Global Envision: May 09, 2007.

\bibitem{46} http://www.alternativabolivariana.org

\end{thebibliography}
\end{document}